\documentclass[aps,twocolumn,showkeys,groupedaddress,prl]{revtex4-1}
\usepackage{graphicx}
\usepackage{amsmath}
\usepackage{amssymb}
\usepackage{longtable}
\usepackage{dcolumn}
\begin{document}

\title{Thermodynamics of water modeled using {\em ab initio} simulations}
\author{Val\'ery Weber}\thanks{Email: valeryweber@hotmail.com}
\affiliation{Physical Chemistry Institute, University of Zurich, 8057 Zurich, Switzerland}
\author{D. Asthagiri}\thanks{Email: dilipa@jhu.edu}
\affiliation{Department of Chemical and Biomolecular Engineering, Johns Hopkins University, Baltimore, MD 21218}

\date{\today}
\begin{abstract}
We regularize the potential distribution framework to calculate the excess free energy of liquid water simulated with the BLYP-D density functional. The calculated free energy is in fair agreement with experiments but the excess internal energy and hence also the excess entropy are not.
Our work emphasizes the importance of thermodynamic characterization in assessing the quality of 
electron density functionals in describing liquid water and hydration phenomena. 
\end{abstract} 


\maketitle

The liquid-vapor coexistence for water is thoroughly characterized experimentally, and
the excess free energy (chemical potential) of a water molecule in the liquid relative to the vapor, $\mu^{\rm ex}_{\rm w}$, a basic descriptor of the liquid-vapor equilibrium, is very well-established. The
chemical potential, together with its derivatives, especially the temperature derivative (the excess entropy), are essential quantities in understanding hydration phenomena and chemical transformations in the liquid state. 

For {\em ab initio} simulations of water,  two earlier studies \cite{asthagiri:pre03,mcgrath:vle06} 
have sought $\mu^{\rm ex}_{\rm w}$.  The first study revealed how the overbinding of water by PW91 (PBE) functionals leads to an overly negative chemical potential and ultimately an over-structured fluid. The 
second explored the coexisting densities of the liquid and vapor phases \cite{mcgrath:vle06} and found a higher than normal vapor density, indicating overbinding of molecules by the BLYP functional. Their limitations notwithstanding, these early studies were incisive in evaluating the fluid simulated by {\em ab initio} dynamics.

In this Letter we present a rigorous calculation of the excess free energy using a theoretical approach that
has matured over the last few years \cite{paliwal:jcp06,shah:jcp07,weber:jcp10}, one that renders calculating free energies from {\em ab initio} simulations far less daunting than before. Together with an independent calculation of the excess internal energy, we obtain the excess entropy as well.  A key finding of this work
is that a satisfactory agreement with experiments of the chemical potential can mask errors in the
excess energy and the entropy, quantities that are crucial in understanding the
thermodynamics of hydration.

The relation between $\mu^{\rm ex}_{\rm w}$ and intermolecular interactions is given by the potential distribution theorem \cite{lrp:book,widom:jpc82}
\begin{eqnarray}
\beta\mu^{\rm ex}_{\rm w} = \ln \int e^{\beta \varepsilon} P(\varepsilon) d\varepsilon \, ,
\label{eq:pdt}
\end{eqnarray}
where $\varepsilon = U_{N} - U_{N-1} - U_{\rm w}$ is the binding energy of the distinguished 
water molecule with the rest of the fluid. $U_{N}$ is the potential energy of the $N$-particle system
at a particular configuration, $U_{N-1}$ is the potential energy of the configuration but with the
distinguished water removed, and $U_{\rm w}$ is the potential energy of the distinguished water
molecule solely. $P(\varepsilon)$ is the probability density distribution of $\varepsilon$ and is
obtained by sampling many configurations of the system. $\mu^{\rm ex}_{\rm w}$ is the
excess free energy in the liquid relative to an ideal gas at the same density and temperature. 
Based on the experimental coexistence densities \cite{wagner:water02} at 298~K, $\mu^{\rm ex}_{\rm w} = -6.3$~kcal/mol.  

A naive application of Eq.~\ref{eq:pdt} to liquid water will fail because the high energy
regions of $P(\varepsilon)$, reflecting the short-range repulsive interactions 
\cite{asthagiri:jacs07,shah:jcp07,asthagiri:jcp2008}, are never well 
sampled in a simulation. We resolve this difficulty by regularizing 
Eq.~\ref{eq:pdt} \cite{asthagiri:jcp2008}. Consider a hard-core solute of radius $\lambda$ centered on a distinguished water molecule: the hard-core solute excludes the remaining water oxygen atoms 
from within the sphere of radius $\lambda$. The chemical potential of the hard-core solute, $\mu^{\rm ex}_{\rm HC}$, is assumed known. With this construction, Eq.~\ref{eq:pdt} 
can be rewritten as \cite{asthagiri:jacs07,shah:jcp07,asthagiri:jcp2008}: 
\begin{eqnarray}
\beta\mu^{\rm ex}_{\rm w} = \beta\mu^{\rm ex}_{\rm HC} & + & \ln x_0 +  \ln \int P(\varepsilon | n_\lambda = 0) e^{\beta \varepsilon} d\varepsilon \, .
\label{eq:mu}
\end{eqnarray}
Here $P(\varepsilon | n_\lambda = 0)$ is the binding energy distribution of the distinguished water molecule 
conditioned on there being no ($n_\lambda = 0$) water oxygen atoms within $\lambda$ of the 
distinguished oxygen atom. By moving the boundary away from the distinguished water, we 
temper the interaction of the distinguished water with the rest of the fluid; indeed, 
for a sufficiently large $\lambda$, $P(\varepsilon | n_\lambda = 0)$ is expected to be well-described by a
Gaussian \cite{shah:jcp07}.  The fraction of configurations that do not have any oxygen atoms within $\lambda$ of the distinguished oxygen is $x_0$; this is the $n=0$ member of the set $\{x_n\}$ of coordination states sampled by the distinguished water molecule and characterizes the interactions between the distinguished water and the contents of the inner-shell. The excess chemical 
potential of the hard-core solute $ \beta \mu^{\rm ex}_{\rm HC} = -\ln p_0$, where $p_0$ is the fraction of configurations in which a cavity of radius $\lambda$ is found in the liquid. $p_0$ is the $n=0$ member of the set $\{p_n\}$ of occupation numbers of a cavity and characterizes the packing interactions in hydration \cite{Pratt:2002p3001}. 

Since $x_0$ and $p_0$ are directly obtained from the simulation record, we only need explicit
energy evaluations to compute the outer-shell contribution
\begin{eqnarray}
\beta \mu^{\rm ex}_{\rm outer} = \ln \int P(\varepsilon | n_\lambda = 0) e^{\beta \varepsilon} d\varepsilon \, . 
\label{eq:invcond}
\end{eqnarray}
Here we calculate the outer-shell contribution using an alternative expression 
\begin{eqnarray}
\beta \mu^{\rm ex}_{\rm outer} = -\ln \int P^{(0)}(\varepsilon | n_\lambda = 0) e^{-\beta \varepsilon} d\varepsilon \, , 
\label{eq:forcond}
\end{eqnarray}
where $P^{(0)}(\varepsilon | n_\lambda = 0)$ is the binding energy distribution in the uncoupled 
ensemble \cite{lrp:book,asthagiri:jcp2008}; that is, 
after we find a suitable cavity in the liquid, we insert a test-particle in that cavity and assess $P^{(0)}(\varepsilon | n_\lambda = 0)$. The superscript $(0)$ indicates that the test-particle and the fluid are thermally uncoupled.  It is advantageous to use Eq.~\ref{eq:forcond} over Eq.~\ref{eq:invcond} because
the uncoupled distribution $P^{(0)}(\varepsilon | n_\lambda = 0)$ has a higher entropy than 
the coupled distribution $P(\varepsilon | n_\lambda = 0)$, rendering the calculation of the free energy of inserting the particle (Eq.~\ref{eq:forcond}) more robust than extracting it (Eq.~\ref{eq:invcond}) \cite{kofke:jcp01}. Further, when $x_0$ is small, it is difficult to characterize $P(\varepsilon | n_\lambda = 0)$ reliabily, a limitation that does not apply to $P^{(0)}(\varepsilon | n_\lambda = 0)$.  

For $\lambda$ sufficiently large such that $P^{(0)}(\varepsilon | n_\lambda =0)$ is Gaussian distributed with a mean $\langle \varepsilon | n_\lambda = 0 \rangle_0$ and variance $\langle \delta\varepsilon^2 | n_\lambda = 0 \rangle_0$ (the subscript 0 emphasizes that the
test-particle and the fluid are thermally uncoupled), Eq.~\ref{eq:forcond} becomes
\begin{eqnarray}
\beta \mu^{\rm ex}_{\rm outer} = \langle \varepsilon | n_\lambda = 0 \rangle_0 - \frac{1}{2k_{\rm B}T} \langle \delta\varepsilon^2 | n_\lambda = 0 \rangle_0 \, .
\label{eq:gaussian}
\end{eqnarray}

We apply the above framework to water simulated with the BLYP-D electron density functional. (In BLYP-D, 
the BLYP  \cite{BLYP:88,DFT_LYP:88} functional is augmented with empirical corrections for dispersion interactions \cite{Grimme:06}.)  We simulate the liquid at a density of 0.997 g/cm$^3$ (number density of 33.33 nm$^{-3}$) and temperature of $350$~K. The higher temperature effectively weakens the bonding \cite{weber:jcp10} and has been found necessary to model the real liquid at the standard density and $T=298$~K \cite{mundy:jpcb09,weber:jcp10}.

We simulate water with the {\sc cp2k} code\cite{cp2knew} and in the $NVE$ and $NVT$ ensembles;
the electronic structure calculations are exactly as in our earlier study \cite{weber:jcp10}, and only salient
differences are noted here. For the studies at $NVT$, we use the hybrid Monte Carlo method  and extended the simulations in our earlier study of a $32$ water system \cite{weber:jcp10}. 
 In the $NVE$ ensemble, we simulated systems with $32$ and $64$ water molecules.

The production phase of the hybrid Monte Carlo lasted 2585 sweeps ($\approx$130~ps). The system temperature was 350~K. Simulations in the $NVE$ ensemble lasted 200~ps of which the last 170~ps were used for analysis. The initial configuration for the 32 particle system was obtained from the end-point of our earlier study \cite{weber:jcp10}. The initial configuration for the 64 water system was obtained from an equilibrated configuration of SPC/E water \cite{spce} molecules. 
In the production phase, the average temperature was $357\pm 26$ K ($362\pm 19$ K) for the 32 (64)
water system. The energy drift was less than 1.8 K (1.2~K) for the 32 (64) water system. 

To calculate $\mu^{\rm ex}_{\rm outer}$, we construct a Cartesian grid within the simulation cell. 
The oxygen population within a radius $\lambda$ of each grid center is calculated. In instances
where there are no ($n_\lambda = 0$) water molecules, we insert a test water molecule in the cavity,
assess its binding energy,  and thus construct $P^{(0)}(\varepsilon|n_\lambda = 0)$. (We draw 
water molecules from the simulation cell, randomly rotate it about an axis through the oxygen center, 
and use the resulting configuration as the test particle.) 

In the $NVE$ simulations, we first find approximately 3500 cavities from  configurations sampled every 50 fs. (The target count of cavities is met by adjusting the spacing of the Cartesian grid between approximately 
0.3 to 0.96~\AA.) Thus over 110k($\approx 32*3500$) binding energy calculations are used to construct $P^{(0)}(\varepsilon|n_\lambda = 0)$.  In the $NVT$ simulations,  the grid spacing for finding cavities to compute $P^{(0)}(\varepsilon|n_\lambda = 0)$ was fixed at 2.0~{\AA}. For $\lambda = 3.0$~{\AA} this gave 98 cavities. In this instance, five different orientations were used for each test particle giving about 16k($\approx 5 * 32 * 98$) binding energies to construct $P^{(0)}(\varepsilon|n_\lambda = 0)$. Subsequently, we resampled the trajectory using a finer grid to better estimate the probability of finding $p_0$.  
Throughout, uncertainties in $x_0$, $p_0$, and $\langle e^{-\beta \varepsilon} | n_\lambda = 0\rangle$ 
were estimated using the Friedberg and Cameron \cite{friedberg:1970} block transformation procedure \cite{allen:error}.

\begin{figure}[t]
\begin{center} 
\includegraphics{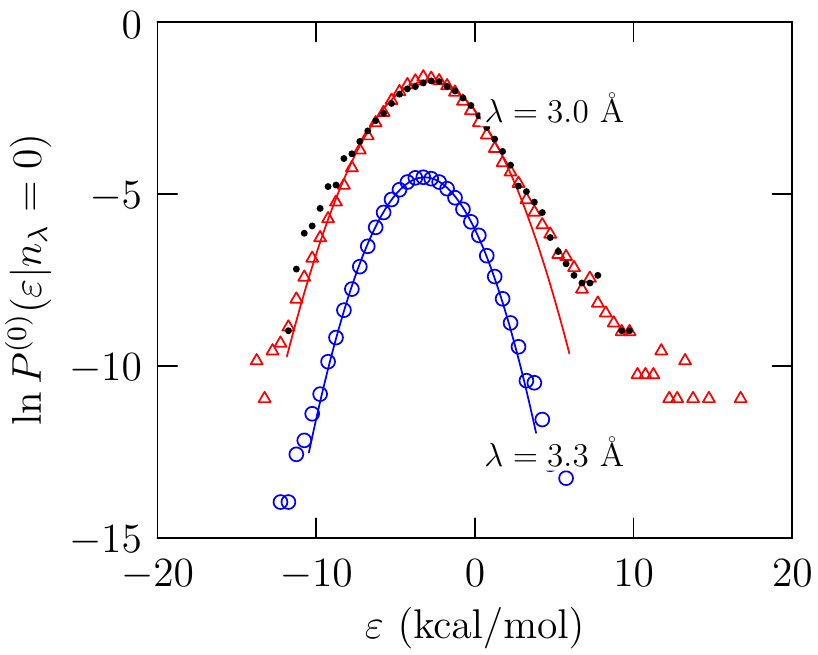}
\end{center}
\caption{\label{fg:sgl_gaus} Distribution of interaction energies $P^{(0)}(\varepsilon|n_\lambda = 0)$ for a 
distinguished water molecule in the center of an empty cavity of radius $\lambda$. The system comprises
32 water molecules. 
The red triangles and blue circles are from the $NVE$ simulation. The solid line defines the
Gaussian model for the respective distribution. The black dots are from the $NVT$ simulation.  The $\lambda = 3.3$~{\AA} curve has been shifted downwards by 3 units for clarity. Observe that 
the thermodynamically important (Eq.~\ref{eq:forcond}) low-energy tail is well-characterized. Further, 
for $\lambda = 3.3$~{\AA}, both wings of the distribution are tempered and the Gaussian model becomes a good  approximation.  }
\end{figure}

Figure~\ref{fg:sgl_gaus} shows $P^{(0)}(\varepsilon|n_\lambda =0)$ for different cavity radii
for the 32 water system. As the figures shows, the low energy region is well-characterized for
inner-shell radii considered here; and it is this low-$\varepsilon$
part of the distribution that is most important for calculating $\mu^{\rm ex}_{\rm outer}$ (Eq.~\ref{eq:forcond}). 

Figure~\ref{fg:sgl_gaus} shows that the agreement between $P^{(0)}(\varepsilon | n_\lambda = 0) $ obtained in the $NVE$ and $NVT$ simulations is satisfactory, although there is somewhat more scatter in the low-energy region for results with $NVT$. This is a consequence of using an order of magnitude less data (16k versus 110k) to construct $P^{(0)}(\varepsilon | n_\lambda = 0)$.   

The probability $x_0$ is a direct measure of the solute-water interactions within the inner-shell, and
perhaps not surprisingly, for large $\lambda$, this quantity is very low. In this case, 
one can model $\{x_n\}$ with a maximum entropy approach \cite{Hummer:1996p326, lrp:jpcb98} using the robust estimates of mean and variance of $\{x_n\}$ and thus secure $x_0$. But for large $\lambda$ \cite{weber:jcp10,paliwal:jcp06}, the underlying assumption of Gaussian occupancy statistics may not be valid. Here, we instead use Bayes's theorem in the form \cite{paliwal:jcp06}
\begin{eqnarray}
x_n=p_{n+1}\frac{p(1|n+1)}{\sum_{m\geq1} p_{m}p(1|m)} \, ,
\label{eq:bayes}
\end{eqnarray} 
where $p(1|m)$ is the conditional probability of finding one water molecule
at the center of the cavity given that $m$ water molecules are present in the cavity. For this purpose, 
following an earlier study \cite{paliwal:jcp06}, the center is any point within 0.15~{\AA} of 
the geometric center of the cavity. 

\begin{figure}[t]
\includegraphics{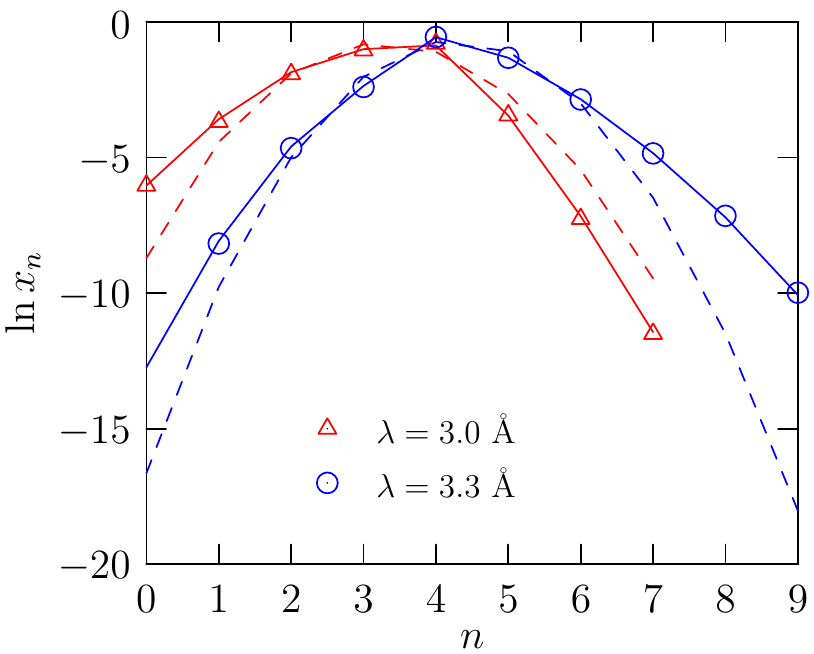}
\caption{\label{fg:xnfit} Observed and estimated $\{x_n\}$ for a $N=64$ water molecule system
and different inner-shell radii $\lambda$. Symbols denote the observed data. The solid line is
obtained using Bayes' theorem (Eq.~\ref{eq:bayes}). The dashed lines are maximum entropy fits
using the mean and variance of $\{x_n\}$ and a flat prior \cite{lrp:jpcb98}. Observe that $\{x_n\}$ obtained using Eq.~\ref{eq:bayes} well-describes the simulation results.}
\end{figure}

Figure~\ref{fg:xnfit} depicts $\{x_n\}$ for the 64 particles system. The fit using Eq.~\ref{eq:bayes} is in excellent agreement 
with the actual data. This gives us confidence in the estimated $x_0$  for $\lambda = 3.3$~{\AA} (Fig.~\ref{fg:xnfit}), an 
instance where $x_0$ was not observed in the simulation. 
As Fig.~\ref{fg:xnfit} shows, the maximum entropy model can lead to errors in $k_{\rm B} T \ln x_0$ 
by about a kcal/mol or more, especially for the large $\lambda$.  

Table~\ref{tb:mu} collects the various components of the excess chemical potential. Across
the range of $\lambda$, the calculated values are within a kcal/mol (often less) of each other and 
the experimental value of $-6.3$ kcal/mol for liquid water at 298 K. Comparing the 32 and 64 
water results shows modest system size effect of about $k_{\rm B} T$. This difference primarily
arises from a less favorable $\mu^{\rm ex}_{\rm outer}$ for the larger system: for a large cavity in a small simulation cell, the outer-shell water molecules pack tightly at the interface and lead to the lower $\mu^{\rm ex}_{\rm outer}$.  
\begin{table*}[t]
  \centering
  \caption{\protect
Chemical  ($k_{\rm B}T\ln x_0$), packing ($k_{\rm B}T \ln p_0$), and long-range interaction ($\mu_{outer}^{ex}$) contributions to the excess chemical potential ($\mu^{\rm ex}_{\rm w}$) of water obtained from $NVE$ molecular dynamics simulations and $NVT$ hybrid Monte Carlo simulations \cite{weber:jcp10}. The radius of the inner shell is 
$\lambda$ (in {\AA}). The average temperatures of the $N=32$ and $N=64$ water systems are, respectively, $357\pm26$~K and $362\pm19$~K. The temperature is 350~K for simulation in the NVT ensemble.  Where available, the inner-shell chemical contribution is evaluated directly from the data, else the estimate using Eq.~\ref{eq:bayes} is used. The outer-shell contribution is calculated by numerical integration of Eq.~\ref{eq:forcond}; estimates from a Gaussian model~\ref{eq:gaussian} (data in parenthesis) are provided for comparison.  Energies  are in kcal/mol. Uncertainties are at 1$\sigma$ level.}\label{tb:mu}
  \begin{ruledtabular}

  \begin{tabular}{ccccccc}
 $N$ &   $\lambda$  & $k_{\rm B}T\ln x_0$ & $k_{\rm B}T\ln x_0$ &   $-k_{\rm B}T\ln p_0$ &           $\mu_{outer}^{ex}$ &  $\mu^{ex}_{\rm w}(l)$ \\
        &                       &        (Simulation) &           (Bayes) &  & & \\
    \hline 
    32 (NVE) & 3.0   & $-4.3\pm0.1$ &  $-4.3$ & $5.5 \pm 0.1$ & $-7.1\pm 0.2$ ($-$6.5) & $-5.9 \pm 0.2$\\

    & 3.1   & $-5.6\pm0.1$ &  $-5.5$ & $6.2 \pm 0.2$ & $-6.5 \pm 0.2$ ($-$6.0) & $-5.9 \pm 0.3$\\
    & 3.2   & $-7.2\pm0.4$ &  $-6.8$ & $7.0 \pm 0.2$ & $-6.2 \pm 0.2$ ($-$5.8) & $-6.4 \pm 0.5$\\
    & 3.3  & $-8.5\pm0.6$ &  $-8.6$ & $8.0 \pm 0.3$ & $-5.7 \pm 0.1$ ($-$5.5) & $-6.2 \pm 0.7$\\ 
(NVT) & 3.0 &  $ -4.4\pm0.1$ &  $ -4.3$ &  $ 5.2\pm0.3$ & $ -7.0\pm0.2 (-7.2)$ &  $-6.2\pm0.4$ \\
    \hline 
    64 (NVE) & 3.0  & $-4.4\pm0.0$ &  $-4.3$  & $5.5 \pm 0.2$ & $-6.8 \pm 0.5$ ($-$5.6) & $-5.7 \pm 0.5$\\
    & 3.1   & $-5.8\pm0.1$ &  $-5.7$ & $6.1 \pm 0.2$ & $-5.8 \pm 0.3$ ($-$5.2) & $-5.5 \pm 0.4$\\
    & 3.2   & $-7.2\pm0.2$ &  $-7.2$ & $6.7 \pm 0.3$ & $-5.3 \pm 0.1$ ($-$4.7) & $-5.8 \pm 0.4$\\
    & 3.3  &       not observed &  $-9.2$  & $7.3 \pm 0.3$ & $-4.8 \pm 0.1$ ($-$4.2) & $-6.7$      \\
  \end{tabular}
  \end{ruledtabular}
\end{table*}
%

We can calculate the excess entropy of hydration per particle from excess chemical potential and the mean binding energy, $\langle \varepsilon\rangle$, of a distinguished water molecule in the fully coupled simulation. First consider the excess internal energy. For the $32$ (64) water system $\langle \varepsilon\rangle = -28 $~kcal/mol ($-27.3$~kcal/mol). Thus the heat of vaporization per particle $\Delta H_{\rm vap} \approx -\langle \varepsilon\rangle / 2 + k_{\rm B}T = 14.6$~kcal/mol ($14.2$~kcal/mol) is substantially in error relative to the experimental value of about $10.5$~kcal/mol at 300~K \cite{wagner:water02}. 

Neglecting small contributions due to the thermal expansion and compressibility of the medium, the excess entropy per particle \cite{shah:jcp07,asthagiri:jcp2008} is $S^{\rm ex}/nk_{\rm B} \approx \langle \varepsilon\rangle / 2k_{\rm B}T - \mu^{\rm ex} / k_{\rm B}T$. Based on the estimates of $\langle \varepsilon\rangle$ and $\mu^{\rm ex}_{\rm w}$ above, it is clear that $T S^{\rm ex}$ for
BLYP-D water is also in error. 

Several important lessons emerge from the present study. First, while BLYP-D at 350~K appears to adequately describe the structure \cite{weber:jcp10} and the hydration free energy relative to experiments, the 
agreement comes due to balancing errors in the excess entropy and excess internal energy. Second, 
while dispersion interactions are regarded as necessary in describing water, the empirical 
dispersion correction in BLYP-D overbinds the material and the temperature of 350~K does 
not adequately compensate for this overbinding.  Finally, the BLYP-D functional at 350~K 
should not be expected to describe the hydration thermodynamics of aqueous solutes, especially 
properties such as the excess entropy of hydration, and thus also transport properties
of solutes in the medium.

\subsection{Acknowledgments}
The authors warmly thank Claude Daul (University of Fribourg) for computer resources. D.A. thanks the donors of the American Chemical Society Petroleum Research Fund for financial support. This research used resources of the National Energy Research Scientific Computing Center, which is supported by the Office of Science of the U.S. Department of Energy under Contract No. DE- AC02-05CH11231.


\end{document}